\documentstyle [11pt]{article}

\begin{document}

\def\beq{\begin{equation}}
\def\eq{\end{equation}}
\def\bar{\begin{eqnarray}}
\def\ear{\end{eqnarray}}
\def\bars{\begin{eqnarray*}}
\def\ears{\end{eqnarray*}}
\def\ov{\overline}
\def\ot{\otimes}

\def\deb{\frac{1}{2}}
\newcommand{\dd}{\mbox{$\Delta$}}
\newcommand{\af}{\mbox{$\alpha$}}
\newcommand{\be}{\mbox{$\beta$}}
\newcommand{\la}{\mbox{$\lambda$}}
\newcommand{\gh}{\mbox{$\gamma$}}
\newcommand{\ep}{\mbox{$\epsilon$}}
\newcommand{\vep}{\mbox{$\varepsilon$}}
\newcommand{\de}{\mbox{$\frac{1}{2}$}}
\newcommand{\th}{\mbox{$\frac{1}{3}$}}
\newcommand{\qa}{\mbox{$\frac{1}{4}$}}
\newcommand{\sx}{\mbox{$\frac{1}{6}$}}
\newcommand{\vg}{\mbox{$\frac{1}{24}$}}
\newcommand{\tde}{\mbox{$\frac{3}{2}$}}

\newcommand{\np}{{\it Nucl. Phys.}}
\newcommand{\pl}{{\it Phys. Lett.}}
\newcommand{\prl}{{\it Phys. Rev. Lett.}}
\newcommand{\cmp}{{\it Commun. Math. Phys.}}
\newcommand{\jmp}{{\it J. Math. Phys.}}
\newcommand{\jpamg}{{\it J. Phys. {\bf A}: Math. Gen.}}
\newcommand{\lmp}{{\it Lett. Math. Phys.}}
\newcommand{\ptp}{{\it Prog. Theor. Phys.}}

\title{$U_q osp(2,2)$ Lattice Models}
\author{\vspace{1cm} Z. Maassarani\\ \small Department of
Physics and Astronomy \\ \small University of
Southern California\\ \small Los Angeles, CA 90089-0484,
USA\thanks{Address starting September $1^{\rm st}$:
Service de Physique Th\'{e}orique, Orme des Merisiers, C. E. --- Saclay,
F-91191 Gif-sur-Yvette Cedex, France.}}
\date{ }
\maketitle
\vspace{1cm}
\centerline{0550; 0590; 1110; 1190}
\begin{abstract}
\begin{normalsize}

In this paper I construct lattice models with an underlying $U_q osp(2,2)$
superalgebra symmetry.
I find new solutions to the graded Yang-Baxter
equation. These {\it trigonometric} $R$-matrices depend on {\it three}
continuous parameters, the spectral parameter, the deformation parameter
$q$ and the $U(1)$ parameter, $b$, of the superalgebra. It must be emphasized
that the parameter $q$ is generic and the parameter $b$ does not
correspond to the `nilpotency' parameter of \cite{gs}. The rational limits
are given; they also depend on the $U(1)$ parameter and this dependence cannot
be rescaled away. I give the Bethe ansatz solution of the lattice models
built from some of these $R$-matrices, while for other matrices, due
to the particular nature of the representation theory of $osp(2,2)$, I
conjecture the result. The parameter $b$ appears as a continuous
generalized spin. Finally I briefly discuss the problem of
finding the ground state of these models.
\end{normalsize}
\end{abstract}
\vspace{2cm}
\hspace{10mm}USC-94/007 \hfill \\
\vspace{8mm}
\hspace{9mm}April 1994 \hfill \\
\vspace{12mm}

\thispagestyle{empty}
\newpage
\setcounter{page}{1}

\section{Introduction}
Lattice models provide us with one way to regularize field theories
by cutting off the short distance divergences. Lattices models
also lead to physical models of statistical and condensed matter
physics (like the Ising model, the Hubbard model, polymer models,...).
Two-dimensional $N=2$ superconformal field theories, which appear in the study
of string theories, are at least as
appealing as lattice models. It is the existence
of a topological sector, for  which a semi-classical approach gives exact
quantum results, that renders $N=2$ theories so attractive. In this sense
$N=2$ theories are simpler than  non-supersymmetric theories.
By studying lattice analogues of $N=2$ models we hope to recover the
`simplified' structure of the field  theory at the lattice level and
to apply $N=2$ supersymmetry to physical models through the lattice
description.

A whole family of lattice analogues of $N=2$
coset models was constructed in \cite{zdn}. The presence
of a topological sector was used to obtain these models.
However the supersymmetry
does not seem to be  realized on the lattice. In an attempt to build
lattice  models through a direct approach and with some degree of
supersymmetry on the lattice, it is natural to consider bosonic
and fermionic lattice
variables. In recent years quantum groups have emerged
as a common underlying symmetry of field theories and lattice models
\cite{bunch}. For instance, $U_{q_+} su(2)\ot U_{q_-} osp(2,2)$ was found
to be an underlying symmetry of $N=2$ superconformal theories, for the
holomorphic part, in the Neveu-Schwarz sector \cite{jime}.

Let $U_q\hat{{\cal G}}$ be the affine $q$-deformed universal enveloping algebra
for a Lie algebra $\cal G$.
A general method for obtaining lattice models, whose states belong
to representations of the Lie algebra, and which have
an underlying $U_q\hat{{\cal G}}$ symmetry, is by making the transfer
matrix out of products of $R$-matrices. The corresponding lattice models
are called vertex models.
Such a method generalizes naturally to superalgebras. In what follows
I consider the $osp(2,2)$ superalgebra.
This is a subalgebra of the $N=2$ superconformal algebra.
In looking for supersymmetry on the lattice,
it is this subalgebra
that one hopes to realize on the lattice.

In this paper I construct lattice models
built from $U_q osp(2,2)$ $R$-matrices. I first obtain new trigonometric
solutions for the Yang-Baxter equation. These solutions depend
on  an additional
continuous parameter $b$ for generic values of $q$. The rational forms
of these matrices also depend on $b$.
(Trigonometric solutions were
constructed in \cite{basha} for the {\em fundamental} representations of simple
Lie superalgebras.) I then
perform an algebraic Bethe ansatz to diagonalize the transfer matrix with
periodic boundary conditions. The parameter $b$ appears as a generalized
spin in  the Bethe ansatz equations. I briefly discuss  the problem of
determining the ground state and the dependence of the  central charge
and conformal weights on the the parameter $b$.

\section{The Superalgebra $osp(2,2)$}

\subsection{Definitions and Notation \label{defnot}}

The $osp(2,2)$ superalgebra is eight-dimensional and has rank two.
There are four even generators, $S_{\pm},S_3$
and $B$, and four odd generators, $V_{\pm}$ and $\overline{V}_{\pm}$.
Let $[\,,]$ denote the usual commutator, and  $\{\,,\}$ denote the
anti-commutator. The commutation relations are given by:

\beq
[S_3,S_{\pm}]=\pm S_{\pm} \; ,\; [S_+,S_-]=2 S_3 \; ,
\label{eq1}
\eq
\beq
[P_{\pm},V_{\pm}]=0 \; ,\; [P_{\mp},V_{\pm}]=\pm V_{\pm} \; ,\;
[P_{\pm},\overline{V}_{\pm}]=\pm \overline{V}_{\pm} \; ,\;
[P_{\mp},\overline{V}_{\pm}]=0 \; ,
\label{eq2}
\eq
\beq
\{V_i,V_j\}=\{\overline{V}_i,\overline{V}_j\}=0 \; ,\; i,j=\pm \; ,
\label{eq3}
\eq
\beq
\{V_+,\overline{V}_-\}=-\de P_+ \; ,\; \{\overline{V}_+,V_-\}=-\de P_- \; ,
\label{eq4}
\eq
\beq
\{V_{\pm},\overline{V}_{\pm}\}=\pm \de S_{\pm} \; ,
\label{eq5}
\eq
where, following the notation of ref. \cite{degu},
\beq
P_{\pm}=S_3 \mp B \; .
\eq
The generators also satisfy the graded Jacobi identity.

The even  sub-algebra is $su(2)\oplus u(1)$, and is generated by
$S_{\pm}$, $S_3$ and $B$, or equivalently, $S_{\pm}$ and $P_{\pm}$.
The generators $V_{\pm}$ (resp. $\overline{V}_{\pm}$) form
$su(2)$ spin-$\de$ tensors (spinors) with `hypercharge'
$B=\de$ (resp. $B=-\de$).

The $osp(2,2)$ superalgebra is a subalgebra of the $N=2$ superconformal
algebra in the NS sector. Using the notation of \cite{nicklec} one has:

\beq
L_0=-S_3 \; ,\;\; L_{\pm}=\pm S_{\pm} \; ,\;\; J_0=2 B \; , \;\;
G_{\pm \frac{1}{2}}^+=2 \overline{V}_{\pm} \; ,\;\;
G_{\pm \deb}^-=2 V_{\pm} \; .
\eq

\subsection{Some $osp(2,2)$ Representations \label{reps}}

Unlike ordinary Lie algebras, the are two types of representations
for most super-algebras. The {\em typical} representations are irreducible and
are similar to the usual  representations of ordinary Lie algebras.
The values of the Casimirs, the central elements, for a given typical
representation, are unique to the  representation.
The {\em atypical} representations have no counterpart
in the ordinary Lie algebra representations. They can be irreducible or not
fully reducible (read reducible but indecomposable).
The Casimirs for two atypical representations can
take the same values.

The superalgebra $osp(2,2)$ is isomorphic to the superalgebra $spl(2,1)$.
The representation theory of the latter superalgebra was studied
in \cite{snr,marcu}.
Generically a representation $(b,s)$
($b \in {\bf C} \; ,\; s=0,\de,1,\tde,...$)
contains four $su(2)\oplus u(1)$ multiplets:
\bar
\label{rep1}
|b,s,s_3\rangle \; &,&\; s_3=-s,-s+1,...,s-1,s \;\; {\rm if}\;\; s\geq 0 \;, \\
|b+\de,s-\de,s_3\rangle \; &,& \;s_3=-s+\de,...,s-\tde,s-\de \;\; {\rm if}\;
\; s\geq \de\;, \\
\label{rep2}
|b-\de,s-\de,s_3\rangle \; &,& \;s_3=-s+\de,...,s-\tde,s-\de \;\;{\rm if}\;\;
s\geq \de\;, \\
\label{rep3}
|b,s-1,s_3\rangle \; &,& \; s_3=-s+1,...,s-2,s-1 \;\;{\rm if}\;\; s\geq 1\; .
\label{rep4}
\ear
The action of the four even generators on these multiplets is the one implied
by the notation. The four odd generators, in contrast, interpolate between
different multiplets. The precise action of these generators, which can be
inferred from the defining eqs. (\ref{eq1}--\ref{eq5}), can be found
in \cite{snr,marcu}. I shall give later the complete matrix form, of
all the generators, for the particular representations I consider.

If $b\neq \pm s$ the representation is denoted by $[b,s]$ and is typical;
the quadratic and
cubic Casimirs do not vanish.
The representation $[b,s]$ has dimension $8s$.

When $b=\pm s$ several kinds of atypical representations arise.
Both Casimirs vanish, and yet these representations are not the
trivial one-dimensional null representation.
One kind has dimension $4s+1$ and is denoted by $[s]_{\pm}$.
To obtain $[s]_+$ (resp. $[s]_-$)
one drops the two multiplets
$(b-\de,s-\de)$ and $(b,s-1)$ (resp. $(b+\de,s-\de)$ and $(b,s-1)$).
These representations are irreducible.
Other kinds of atypical representations, for which all multiplets are kept,
have dimension $8s$.
They contain two representations of the
previous type with one representation being an invariant subspace of the
whole representation. They are therefore not fully reducible.

Other types of atypical representations exist, with different
dimensionalities.

\section{$U_q osp(2,2)$}

I show how the universal enveloping algebra $U osp(2,2)$ is deformed.
The coproduct is introduced in order to define the universal, spectral
parameter-independent $R$-matrix, which depends on $q$.
This matrix will be useful in obtaining spectral parameter-dependent
$R$-matrices, $R(x,q)$, which give integrable lattice models.
The matrix elements of the matrix $R(x,q)$ give the Boltzmann
weights of an integrable lattice model.
The coproduct will also be used to obtain another spectral parameter-dependent
$R$-matrix, which is one ingredient in the Bethe ansatz diagonalization.

\subsection{The q-deformed Relations}

I consider the $q$-deformation $U_q osp(2,2)$ of the universal enveloping
algebra, $U osp(2,2)$, obtained
in ref. \cite{degu}. Another deformation  exists in the literature \cite{defo}.
The latter deformation relies on a harmonic oscillator representation
of the superalgebra.

The $q$-deformation considered here is precisely the one
that appears as an underlying quantum group symmetry of $N=2$ superconformal
theories in the Neveu-Schwarz sector \cite{jime}.
There is also another reason for this choice.
Superalgebras, unlike ordinary Lie algebras, admit, in general, more
than one inequivalent basis of simple roots. It turns out
that one can construct $N=2$  supersymmetric Toda field theories
only if one chooses a purely fermionic simple root system \cite{hollow}.
This should be taken
as an additional hint if one has in mind the construction of a supersymmetric
theory. The foregoing deformation is based on the choice of a purely
fermionic simple root system.

The universal enveloping algebra $U osp(2,2)$ is generated by $P_{\pm},
V_{\pm}, \overline{V}_{\pm}$. This corresponds to an implicit choice of two
purely fermionic simple roots for the basis. The relations (\ref{eq2}),
(\ref{eq3}) and
(\ref{eq5}) are kept unchanged except for the fact that the generators
are the $q$-deformed ones. The two relations (\ref{eq4}) are replaced by
\beq
\{2 V_+,2 \overline{V}_-\}= [-2 P_+] \; ,\;\;
\{2 \overline{V}_+, 2 V_-\}=[-2 P_-] \; ,
\label{eq6}
\eq
where
\beq
[x] \equiv \frac{q^x-q^{-x}}{q-q^{-1}} \; .
\label{qana}
\eq

The $q$-deformations of relations (\ref{eq1}) are obtained from (\ref{eq5})
and (\ref{eq6}).
The $q$-generators $S_{\pm}, S_z$ do {\it not} satisfy
the usual $q$-deformed $su(2)$ relations:
\beq
[S_3,S_{\pm}]=\pm S_{\pm}\;,\; [S_+,S_-]= [2S_3] \;.
\eq
It is easy to verify that the
first relation in (\ref{eq1}) remains the same while the second
relation  becomes:
\bar
[S_+,S_-]=-V_- \ov{V}_+ (q^{2 P_+ +1}+q^{-2 P_+ -1})
-\ov{V}_+ V_- (q^{2 P_+ -1}+q^{-2 P_+ +1}) \nonumber \\
-\ov{V}_- V_+ (q^{2 P_- +1}+q^{-2 P_- -1})
-V_+ \ov{V}_- (q^{2 P_- -1}+q^{-2 P_- +1}) \; .
\ear
This relation collapses to the usual one, $[S_+,S_-]= [2S_3]$,
for certain representations.
It is interesting to note that the $su(2)$ subalgebra is not deformed
to a $U_q su(2)$ subalgebra of $U_q osp(2,2)$.
This seems to be the case for superalgebras in general
(see \cite{salosp} for example).

\subsection{Coproducts}

The usual tensor product construction, {\em for the operators}, of the algebra
is not compatible with the
$q$-deformed commutation relations of the (Hopf) superalgebra $U_q osp(2,2)$.
However a $q$-deformed tensor product can be defined. This new tensor product
is conveniently encoded in the coproduct $\Delta$.
The defining relations are given by \cite{degu}:
\bar
\label{deltav0}
\dd (P_{\pm}) = P_{\pm}\ot 1 + 1\ot P_{\pm} \; , \\
\dd (V_{\pm}) = q^{P_{\pm}}\ot V_{\pm} + V_{\pm}\ot q^{-P_{\pm}} &,&
\dd (\ov{V}_{\pm}) = q^{P_{\mp}}\ot \ov{V}_{\pm} + \ov{V}_{\pm}\ot
q^{-P_{\mp}} \; .
\label{deltav}
\ear
Again $\dd (S_{\pm})$ are not given by the usual $U_q su(2)$ expressions;
they can be obtained using (\ref{eq5}) and (\ref{deltav}).

Throughout this chapter the explicit tensor product sign $\ot$
is graded. This means that the following rule is
applied: minus signs are generated each time two odd elements, generators
and/or vectors (in some representation) are
`commuted' through one another.

There is also another coproduct, $\ov{\Delta}$, with defining
relations obtained by pairwise permuting the generators in the defining
relations of $\Delta$ ({\it i.e. }$A\ot B \rightarrow B\ot A$). Equivalently,
one has $\ov{\Delta}_q =\Delta_{q^{-1}}$, for the explicit dependence on $q$ in
the defining relations (\ref{deltav}).
However, as happens with $q$-deformations of ordinary Lie algebras,
it is possible to show that there exists a `matrix' $R$ \cite{degu},
in $U_q osp(2,2) \ot U_q osp(2,2)$, such that
the coproducts $\Delta$ and $\ov{\Delta}$ are related by
\beq
\ov{\Delta}(g)R=R\Delta(g)
\label{rd}
\eq
for all elements $g$ of $U_q osp(2,2)$.
The `matrix' $R$ is universal because it is representation independent
and depends only on the algebra.

The matrix $R$ satisfies the Yang-Baxter equation
\beq
R_{1 2} R_{1 3} R_{2 3} = R_{2 3} R_{1 3} R_{1 2}
\label{ybeq}
\eq
where, for $R=\sum a\ot b$, $R_{1 2}=\sum a\ot b\ot 1$, $R_{1 3}= \sum
a \ot 1 \ot b$, etc...

It is the extension of this construction to affine (Kac-Moody)
$q$-deformed algebras in general which gives the $R$-matrix an additional
dependence on the spectral parameter $x$.

\section{A New $R$-matrix}

I now determine new $R$-matrices for a set of four-dimensional
representations of $osp(2,2)$. The fundamental representation, $[0,\de]$, of
the supergroup $OSp(2,2)$ is contained in the set of four-dimensional
representations $[b,\de]$.

I first define and then $q$-deform the representations $[b,\de]$.
I obtain the matrix $R(q,b)$ and its spectral decomposition, and then find
the `affinized', {\em i.e.} spectral parameter dependent, version of the
matrix $R(q,b)$.

\subsection{The $[b,\de]$ representations}

I consider the four-dimensional typical representations
$[b,\de]$, in the notation of
section (\ref{reps}), where, {\it a priori}, $b\in {\bf C}-\{\pm\de\}$.
One has inequivalent representations for
different values of $b$. From (\ref{rep1}--\ref{rep4}), one finds four vectors
which I choose as a basis:
\beq
b_1=|b,\de\rangle \; ,\; f_1=|b-\de,0\rangle \; ,\; f_2=|b+\de,0\rangle \; ,
\; b_2=|b,-\de\rangle \; ,
\label{basis}
\eq
where $b_1$ and $b_2$ are even (bosonic) and $f_1$ and $f_2$ are
odd (fermionic). The {\em relative} parities, two even and two odd,
of those vectors can be chosen without loss of generality. The two resulting
$R$-matrices are related by a similarity (gauge) transformation.

The generators for the representation $[b,\de]$ are given by \cite{snr,marcu}:
\bar
\label{vrep1}
V_+=\left( \begin{array}{cccc}
0 & \ep & 0 & 0 \\
0 & 0 & 0 & 0 \\
0 & 0 & 0 & \af \\
0 & 0 & 0 & 0    \end{array} \right) &,&
\ov{V}_-=\left( \begin{array}{cccc}
0 & 0 & 0 & 0 \\
-\be & 0 & 0 & 0 \\
0 & 0 & 0 & 0 \\
0 & 0 & \gh & 0  \end{array} \right)\;, \\
\label{vrep2}
\ov{V}_+=\left( \begin{array}{cccc}
0 & 0 & \gh & 0 \\
0 & 0 & 0 & \be \\
0 & 0 & 0 & 0 \\
0 & 0 & 0 & 0  \end{array} \right) &,&
V_-=\left( \begin{array}{cccc}
0 & 0 & 0 & 0 \\
0 & 0 & 0 & 0 \\
-\af & 0 & 0 & 0 \\
0 & \ep & 0 & 0  \end{array} \right) \;, \\
P_+=\left( \begin{array}{cccc}
\de -b & 0 & 0 & 0 \\
0 & \de -b  & 0 & 0 \\
0 & 0 & -\de -b  & 0 \\
0 & 0 & 0 & -\de -b  \end{array} \right)\!\! &,& \!\!
P_-=\left( \begin{array}{cccc}
b + \de & 0 & 0 & 0 \\
0 & b- \de & 0 & 0 \\
0 & 0 & b + \de & 0 \\
0 & 0 & 0 & b -\de \end{array} \right)\,. \nonumber
\ear
The four parameters appearing in (\ref{vrep1}--\ref{vrep2}) are constrained
by
\beq
4 \af \gh =1+2 b \;\; ,\;\;\; 4 \be \ep = 1 - 2 b \; .
\label{const}
\eq
Thus there are two free parameters which correspond to arbitrary relative
normalizations of the  $su(2)$ doublet $(b_1,b_2)$,
and the two singlets $f_1$ and $f_2$.

For $b\neq \de$
these representations are typical while, for $b=\pm \de$ they are atypical.
Note that $b$ can take complex values
in what follows if one does not consider hermitian representations.

The $q$-deformed version of this representation is obtained as follows.
The generators $P_{\pm}$ remain undeformed. The matrix form
of the four odd generators remains unchanged. However the four parameters
satisfy $q$-deformed constraints:
\beq
4 \af \gh =[1+2 b] \;\; ,\;\;\; 4 \be \ep = [1 - 2 b] \; ,
\label{qconst}
\eq
where $[x]$ was defined in (\ref{qana}).
If $q$ goes to one then equations (\ref{qconst}) yield equations
(\ref{const}) and one recovers the undeformed representation.

\subsection{The Matrix $R(q,b)$}

The matrix $R(q,b)$ is obtained by plugging the four-dimensional
matrices, (\ref{vrep1}--\ref{vrep2}) and $P_{\pm}$,
subject to the deformation constraints
(\ref{qconst}), into the universal $R$-matrix found in \cite{degu}.
I shall give the result in a slightly different form below.

I now put the $R$-matrix just obtained in a form which is convenient to give
it a spectral parameter dependence. Recall that $R(q,b)$ acts in $V \ot V$.
Because the two
representations on the left and the right of this tensor product are equal,
one can define a, graded, permutation operator, $\cal P$. It satisfies
\beq
{\cal P} |u\rangle \ot |v\rangle = (-1)^{\vep_u \vep_v}|v\rangle \ot |u\rangle
\eq
for two vectors, $u$ and $v$, of definite parity, $\vep_u$ and  $\vep_v$,
in the representation space $V$.
One can then define
the matrix $\check{R}$ given by:
\beq
\check{R}={\cal P} R\; .
\eq
{}From (\ref{rd}) one gets
\beq
\check{R}\dd(g) =\dd(g)\check{R}\;,
\eq
for all $g \in U_q osp(2,2)$.
This means that $\check{R}$ commutes with the action of $U_q osp(2,2)$,
and therefore one has a decomposition in terms of projectors.

The tensor product of two representations
$[b,\de]$ has the form:
\beq
[b,\de]\otimes [b,\de]= [2 b,1]\oplus [2 b +\de,\de] \oplus [2 b -\de,\de] \; ,
\label{tprod}
\eq
with dimensionalities 8, 4 and 4 for the right-hand-side.
When $b=0$, the two four-dimensional representations coalesce and form one
eight-dimensional atypical representation (see refs. \cite{snr,marcu}).
For this reason, the case $b=0$ should be, and is, treated with some care.

I have found the eigenvalues and eigenvectors of $\check{R}(q,b)$,
and then obtained the projectors form these
eigenvectors.
The $\check{R}$-matrix can be written as
\beq
\check{R}(q,b)=q^{1-4 b^2} P_1(q,b) -q^{-(1+2 b)^2} P_2(q,b)
- q^{-(1-2 b)^2} P_3(q,b) \; ,
\label{spec}
\eq
where the $P_i$'s form a complete set of orthogonal projectors, {\em i.e.}
$P_i P_j =\delta_{ij} P_i$.
They correspond to $[2 b,1]$ ($P_1$), $[2 b +\de,\de]$ ($P_2$)
and $[2 b -\de,\de]$ ($P_3$). For $b=0$, one takes a limit where the
projectors $P_2$ and $P_3$ are combined.
Define\footnote{The symbol $[s]_{\pm}$ was
also used in section (\ref{reps}) for atypical representations. However
there is no room for confusion.}
\beq
[x]_{\pm}\equiv q^x \pm q^{-x}\; .
\eq
I give below the non-vanishing elements of the projectors, which have
a block-diagonal structure. One has

(i) one-dimensional sub-matrices,
\bar
P_1 &=&1 \; ,\; P_2=P_3=0 \;\; {\rm for} \;\; b_i\otimes b_i \; ,
\; i=1,2 \; , \\
P_2 &=&1 \; ,\; P_1=P_3=0 \;\; {\rm for} \;\; f_2\otimes f_2 \; , \\
P_3 &=&1 \; ,\; P_1=P_2=0 \;\; {\rm for} \;\; f_1\otimes f_1 \; ,
\ear

(ii) two-dimensional sub-matrices,
\beq
P_1=\frac{1}{[1-2b]_+} \left(
\begin{array}{cc}
q^{1-2b} & 1 \\
1 & q^{-1+2b} \end{array} \right) \; ,\;
P_3=\frac{1}{[1-2b]_+} \left(
\begin{array}{cc}
q^{-1+2b} & -1 \\
-1 & q^{1-2b} \end{array} \right) \; ,\;
P_2=0 \; ,
\eq
for the two bases $(b_1\otimes f_1, f_1\otimes b_1)$ and
$(f_1\otimes b_2, b_2\otimes f_1)$,
\beq
P_1=\frac{1}{[1+2b]_+} \left(
\begin{array}{cc}
q^{1+2b} & 1 \\
1 & q^{-1-2b} \end{array} \right) \; ,\;
P_2=\frac{1}{[1+2b]_+} \left(
\begin{array}{cc}
q^{-1-2b} & -1 \\
-1 & q^{1+2b} \end{array} \right) \; ,\;
P_3=0 \; ,
\eq
for the two bases $(b_1\otimes f_2, f_2\otimes b_1)$
and $(f_2\otimes b_2,b_2\otimes f_2)$,

(iii) four-dimensional sub-matrices,
\bars
P_2(q,b) &=&\frac{1}{D} \left(
\begin{array}{cccc}
q^{-4b} & q^{-2b} \frac{[1-2b]_-^{1/2}}{[1+2b]_-^{1/2}} &
q^{-2b} \frac{[1-2b]_-^{1/2}}{[1+2b]_-^{1/2}} & -1 \\
-q^{-2b} \frac{[1-2b]_-^{1/2}}{[1+2b]_-^{1/2}} & -\frac{[1-2b]}{[1+2b]} &
-\frac{[1-2b]}{[1+2b]} & q^{2b} \frac{[1-2b]_-^{1/2}}{[1+2b]_-^{1/2}} \\
-q^{-2b} \frac{[1-2b]_-^{1/2}}{[1+2b]_-^{1/2}} & -\frac{[1-2b]}{[1+2b]} &
-\frac{[1-2b]}{[1+2b]} & q^{2b} \frac{[1-2b]_-^{1/2}}{[1+2b]_-^{1/2}} \\
-1 & -q^{2b} \frac{[1-2b]_-^{1/2}}{[1+2b]_-^{1/2}} &
-q^{2b} \frac{[1-2b]_-^{1/2}}{[1+2b]_-^{1/2}} & q^{4b}
\end{array} \right) \; , \\
P_3(q,b) &=& P_2(q,-b) \; ,\; P_1(q,b)=Id_4-P_2(q,b)-P_3(q,b) \; ,
\ears
where $D=[4b]_+ -2\frac{[1-2b]}{[1+2b]}$,
for the basis $(b_1\otimes b_2,f_1\otimes f_2,f_2\otimes f_1,b_2\otimes b_1)$.
The four-dimensional pieces of the projectors depend, {\em a priori},
on $\af$, $\be$,
$\gh$ and $\ep$. I have chosen
$\af=\gh$ and $\be=\ep$, which I call the symmetric choice because
$\check{R}$ is a symmetric matrix, except for some minus signs in the
four-dimensional pieces.
Other choices for $\af$, $\be$, $\gh$ and $\ep$, yield $R$-matrices that
are gauge-equivalent to the foregoing matrix.

All the foregoing analysis was done using a graded tensor product. The
scalar product on the the space $[b,\de]\otimes [b,\de]$ is the graded induced
scalar product, which is therefore not positive definite.

\subsection{The Matrix $\check{R}(x,q,b)$}

To construct a two-dimensional lattice model one needs the spectral parameter
dependent $R$-matrix, or the `baxterized' form of the foregoing matrix.
I find:
\beq
\check{R}_{\la}(x,q,b)=\frac{\la x q^{4 b^2 -1}+1}{x+\la q^{4 b^2 -1}}
P_1(q,b) + \frac{-\la x q^{(1+2 b)^2} +1}{x-\la q^{(1+2 b)^2}} P_2(q,b)
+\frac{-\la x q^{(1-2 b)^2} +1}{x-\la q^{(1-2 b)^2}} P_3(q,b) \; .
\label{rx}
\eq
The spectral parameter $x$ is a multiplicative one (see (\ref{ybe})).
The discrete `index' $\la$ in (\ref{rx}) can take the value
$q^{1-4 b^2}$ for {\em all} $b$, or the value $-q^{-1}$ for $b=0$.
Thus (\ref{rx})
consists of two different matrices. I think of (\ref{rx}) as defining one
matrix, and comment on the
appearance of $\la$ in the following section.
The spectral parameter dependent $R$-matrix is obtained as $R(x,q,b)=
{\cal P}\check{R}(x,q,b)$. It satisfies the graded Yang-Baxter
equation\footnote{These calculations and others were carried out using
{\em Mathematica}$^{\rm TM}$.}
\beq
R_{1 2}(xy^{-1},q,b)R_{1 3}(x,q,b)R_{2 3}(y,q,b)=
R_{2 3}(y,q,b)R_{1 3}(x,q,b)R_{1 2}(xy^{-1},q,b) \; .
\label{ybe}
\eq
The $R$-matrices considered here are even operators. However they are given
by linear combinations of tensor products of odd and even operators, and
therefore $R_{1 3}$ does not act trivially on the second space (as the
subscripts would imply for the non-graded case). Hence
the Yang-Baxter equation is, in components,
\beq
R_{i_1 i_2,j_1 j_2} R_{j_1 i_3,l_1 k_3} R_{j_2 k_3, l_2 l_3}
(-1)^{\varepsilon_{j_2}(\varepsilon_{i_3} + \varepsilon_{k_3})}
=R_{i_2 i_3,j_2 j_3} R_{i_1 j_3,k_1 l_3} R_{k_1 j_2, l_1 l_2}
(-1)^{\varepsilon_{j_2}(\varepsilon_{j_3} + \varepsilon_{l_3})} \; ,
\label{ybec}
\eq
where $\varepsilon_i = 0,1$. The arguments of the matrices are the same as in
eq. (\ref{ybe}).
The matrix $\tilde{R} \equiv (-1)^{\varepsilon_{i_1} \varepsilon_{i_2}}
R_{i_1 i_2,j_1 j_2}$
satisfies the ordinary Yang-Baxter equation.
Therefore modulo some redefinitions the grading signs can be removed;
however this does not bring any real simplifications.
For $\la = q^{1-4 b^2}$ the matrix $R(x,q,b)$ also satisfies equations
(\ref{aff1}--\ref{aff}), where the representations on the left and right of the
tensor product sign $\otimes$ are both equal to $[b,\frac{1}{2}]$.

\subsection{Comments}

The matrix (\ref{rx}) has some unusual and interesting properties.
This matrix is a {\it trigonometric}
$R$-matrix
which depends on {\it three} continuous and arbitrary complex parameters,
$x$, $q$ and $b$.
This seems to be a new result \cite{bgzd}.
Usually a three-parameter dependence is associated with
{\it elliptic} solutions of the Yang-Baxter equation. In this respect
the free fermion model, which depends on three complex variables, enters the
context of elliptic $R$-matrices \cite{bastro}.
Trigonometric $R$-matrices with
three parameters, for $U_q su(2)$, were found
in refs. \cite{gs}. However, these matrices only exist at roots of
unity ($q^N=1$), and correspond to {\em nilpotent} irreducible
representations.

The dependence on the parameter
$b$, a representation label, is immediately traced back to
the non-compact generator $B$ of the
$osp(2,2)$ superalgebra. One can also obtain $R$-matrices which depend
on an additional parameter, $b'$, by  considering an $R$-matrix for the
representation
$[b,\de]\otimes [b',\de]$, or for higher dimensional representations.

One can, presumably, construct
trigonometric solutions that depend on a higher number of continuous
parameters,
if the algebra has a higher number of non-compact generators, and/or
one considers representations at roots of unity.

The two matrices obtained for the
two values  $\la=q$ (for $b=0$) and $\la=-q^{-1}$ (for $b=0$),
are {\it not} related by a gauge transformation. A gauge transformation
is a diagonal similarity transformation; it preserves the Yang-Baxter
equation. This `doubling' occurs here at the {\it affine} level.
The existence of two inequivalent
$R$-matrices for the same representation was noticed for another rank two
algebra, $U_q su(3)$ \cite{su3}, not $U_q \widehat{su(3)}$.
However here, the two matrices
may correspond to two inequivalent coproducts for the
affine superalgebra $U_q\widehat{osp(2,2)}$.

The $R$-matrix for the fundamental representation $[b=0,\de]$ of $OSp(2,2)$
was given in ref. \cite{degu}. One can recover it by taking
$\la=-q^{-1}$ and the limit $b=0$.\footnote{This matrix does not however
satisfy (\ref{aff}).} The tensor product of the
two representations $[0,\de]$ gives two eight-dimensional representations (as
noted earlier).
One of these representations is atypical. The limit does exist, as can
be seen by carefully considering the projectors $P_2$ and $P_3$: they combine
as expected to form an  eight-dimensional projector onto the atypical
representation, and a singlet `projector' for the one-dimensional invariant
subspace in it. The limit also exists for the other value of $\la$.

Similarly, the limits $b=\pm \de$ are not straightforward. One can make sense
of these limits from the final form (\ref{rx}). However, because (\ref{rx})
corresponds to the
the `symmetric choice', the limit corresponds to the tensor product
of two direct sums $[0]\oplus [\de]_{\pm}$.

I give for completeness, in appendix, the rational
limits obtained from (\ref{rx}). Rational limits can be obtained by letting
both $x$ and $q$ tend to one, with $x$ behaving as some power of $q$.

I have obtained new trigonometric $R$-matrices which depend on three
parameters.
I believe these matrices have features which were not found before
in the literature.

\section{Bethe Ansatz}

\subsection{The Need for Fusion}

The matrix (\ref{rx}) has exactly thirty six non-vanishing matrix elements.
The corresponding 36-vertex model cannot be solved using a direct Bethe
ansatz approach. There are simply too many non-vanishing matrix elements.
This is due to the fact that the representation $[b,\de]$ is not the
smallest representation of the algebra $osp(2,2)$. Because of the `excess'
of matrix elements, the algebraic $RTT$ relations, which play a central
role in the algebraic Bethe ansatz method, do not have the form which
allow a Bethe ansatz. For this same reason too, the obvious candidates for
`highest weights vectors', from which the Bethe ansatz eigenvector is built by
applying lowering operators, do not have the required properties. Namely,
there are too many elements of the monodromy matrix, (\ref{mono}), which do not
annihilate these candidates. The situation is similar for higher dimensional
representations of $q$-deformed Lie algebra. A fusion procedure is used for
all those cases. See ref. \cite{resh} for $U_q su(2)$ for instance.

In what follows I take $\lambda = q^{1-4b^2}$ (in (\ref{rx})).

\subsection{Auxiliary and Quantum Spaces}

Fusion requires involving other lattice models. To do this, one
finds solutions of
the Yang-Baxter equation (the
dependence on the other parameters is implicit)
\beq
R_{1 2}(x y^{-1})R_{1 3}(x)R_{2 3}(y) =
R_{2 3}(y)R_{1 3}(x)R_{1 2}(x y^{-1}) \; ,
\label{mix}
\eq
where the three spaces involved are not necessarily copies of the same space.
The matrices $R$ exist
for the tensor product of any representations of a given $q$-deformed Lie
algebra, or superalgebra. The computations needed to find these matrices are
not usually simple. Now recall how one obtains
transfer matrices of vertex models.
Define then the local operator $L_n(x)$, at site $n$ and for an auxiliary
space denoted by `a', by
\beq
L_n(x)=R_{{\rm a}n}(x) \; .
\label{l}
\eq
The corresponding monodromy matrix, $T(x)$, is
\beq
T(x)= L_L(x)\otimes L_{L-1}(x)\otimes ... \otimes L_1(x) \; ,
\label{mono}
\eq
where the tensor product is graded, and $L$ is the number of sites.
The transfer matrix is given by
\beq
\tau (x)= {\rm Str}_{\rm a}(T(x))\equiv\sum_a (-1)^{\varepsilon_a}T_{aa}(x)\;,
\eq
where ${\rm Str}_{\rm a}$ is the supertrace over the auxiliary space `a'.
I am considering here periodic boundary conditions. I
shall come back to this point later.

Equations (\ref{mix}), (\ref{l}) and (\ref{mono}) imply that the
monodromy matrices intertwine according to:
\beq
R_{1 2}(x y^{-1}) \stackrel{1}{T}(x) \stackrel{2}{T}(y) =
\stackrel{2}{T}(y)\stackrel{1}{T}(x) R_{1 2}(x y^{-1}) \; ,
\label{rtt}
\eq
where $\stackrel{1}{T}(x)=T(x)\otimes 1$ and $\stackrel{2}{T}(y)=1\otimes
T(y)$.

The two auxiliary spaces, 1 and 2 in the subscripts and `top-scripts'
of (\ref{rtt}), are not necessarily the same.
Both transfer matrices act in the same quantum space, which is the space
3 in (\ref{mix}) tensored $L$ times.
The supertrace over the two auxiliary spaces  1 and 2, of the  $RTT$
relations (\ref{rtt}), gives the commutations of the transfer matrices
at different spectral parameters:
\beq
[\tau_1 (x),\tau_2 (y)] = 0 \; .
\label{com4}
\eq
The subscripts in (\ref{com4}) are just to remind us that the
two auxiliary spaces
can be different. This commutation relation implies that the transfer
matrices can be diagonalized {\it simultaneously}.

\subsection{Fusion}

The second ingredient in the fusion procedure consists of obtaining
relations between some transfer matrices that commute, as in (\ref{com4}).
An $R$-matrix `degenerates', or becomes block-diagonally proportional
to projectors, at
certain values of the spectral parameter $x$. The projections are
on representations which appear in the tensor product of the representations,
for which the $R$-matrix was built. This degeneracy is a generic feature
of $R$-matrices.

The projectors are used to construct $R$-matrices for the representations
corresponding to the projectors. I give explicit examples of such
a degeneracy and fusion in the following sections.

\subsection{Some Technical Considerations \label{philo}}

I follow the philosophy of ref. \cite{resh}. One wants to
diagonalize, by algebraic Bethe ansatz, the transfer matrix
of the model constructed
from the $R$-matrix of $V_1\ot V_2$, for two representations $V_1$ and $V_2$
of a certain Lie algebra (or superalgebra). If $V_1$, the auxiliary
space, is the `fundamental' representation of the algebra
then a diagonalization
by algebraic Bethe ansatz can be performed without further ado. This
is because the monodromy matrices intertwine with the $R$-matrix of
the tensor product of two fundamental representations.
If $V_1$ is not `the' fundamental, it should obtained by tensor product
of the fundamental and smaller representations. The Bethe ansatz is performed
for $({\rm fund.\; rep.})\ot V_2$. The
eigenvalues, for the transfer matrix of the lattice model $V_1\ot V_2$,
are obtained
using the fusion equations and the commutation of the intermediate
transfer matrices.

Note that higher representations can be considered. They give relations
between transfer matrices, but are not really helpful for doing a
Bethe ansatz.

The $osp(2,2)$ algebra has two three-dimensional representations, $[\de]_+$
and $[\de]_-$. These are the smallest non-trivial representations. I
consider $[\de]_+$.
As we shall see, the $R$-matrix of $[\de]_+\ot [\de]_+$ has,
a form which allows
an algebraic Bethe ansatz. To `bridge the gap' between
$[\de]_+$ and $[b,\de]$, I determine the $R$-matrix for
$[\de]_+\ot [b,\de]$. The monodromy matrices of this tensor product intertwine
using the transfer matrix of $[\de]_+\ot [\de]_+$. An algebraic
Bethe ansatz for the transfer matrix of $[\de]_+\ot [b,\de]$ is possible.
I do this in section (\ref{bean}). Then, to remain faithful to the foregoing
philosophy, one would like to obtain the representation
$[b,\de]$ in a tensor product of representations $[\de]_{\pm}$. However,
after careful consideration, the only relevant tensor product is of
$[\de]_+$ with itself:
\beq
[\de]_+\otimes [\de]_+ = [\tde,\de] \oplus [1]_+ \; ,
\eq
with dimensions 4 and 5 for the right-hand side.

The representation theory of $osp(2,2)$ is such that, of all the
four-dimensional representations $[b,\de]$, only $[\pm\tde,\de]$ can be
obtained in the tensor product of the three-dimensional representations.
In this sense, the representations $[b,\de]$ are `fundamental' too.

Nevertheless, a fusion in the auxiliary space to obtain the transfer matrix
of $[\tde,\de]\ot [b,\de]$, the eigenvalues, and the Bethe ansatz equations,
is useful. I use these results to conjecture the  eigenvalues and
Bethe ansatz equations of the $R$-matrix of $[b,\de]\ot [b,\de]$.

Let $3=[\de]_+$, $4'=[\tde,\de]$ and $4=[b,\de]$ to simplify the notation.
One then has five $R$-matrices
to  consider, $R^{3\otimes 3}$ , $R^{3\otimes 4'}$, $R^{3\otimes 4}$,
$R^{4'\otimes 4}$ and $R^{4\otimes 4}$.

In what follows the spectral parameters are arbitrary.
For the first and the second space equal to the
three-dimensional representation, and the third space equal to $[b,\de]$, in
(\ref{mix}),
one obtains a monodromy matrix with a $[\de]_+$-auxiliary space and a
$[b,\de]$-quantum space.
These $T^{(3 4)}$ monodromy matrices intertwine
as in (\ref{rtt}) with the matrix $R^{3\ot 3}$,
and therefore their transfer matrices,
which act in the tensor products of $[b,\de]$-spaces,
commute. It is these transfer matrices that are directly diagonalized
by Bethe ansatz.
A first space equal to $[\de]_+$, a second space equal to $[\frac{3}{2},
\de]$, and a third space equal to $[b,\de]$, give monodromy matrices
$T^{(34)}$ and $T^{(4'4)}$ intertwining
according to a $[\de]_+\otimes [\frac{3}{2},\de]$-matrix. The transfer matrices
again commute. A fusion equation (\ref{fus1}) is then found. It relates the
relevant eigenvalues.
Similarly, considering $(1,2,3)=(4',4,4)$, in an obvious notation,
ensures the possibility of
simultaneous diagonalization of the $\tau^{(4'4)}$ and $\tau^{(44)}$ matrices.
Considering $(1,2,3)=(4',4',4)$ ensures the possibility of
simultaneous diagonalization of $\tau^{(4'4)}$ matrices.
Finally, considering $(1,2,3)=(4,4,4)$ ensures the possibility of
simultaneous diagonalization of $\tau^{(44)}$ matrices.

\subsection{The $R$-matrix for $[\de]_+\ot [\de]_+$ \label{r33}}

Let $B_1=|\de,\de\rangle$, $F=|1,0\rangle$ and $B_2=|\de,-\de\rangle $
form the basis of $[\de]_+$. These vectors should not be confused with
the four vectors defined earlier. The $q$-deformed generators are
\bar
P_+={\rm diag}(0,-1,-1) \; ,\; P_-={\rm diag}(1,1,0) \; , \\
(V_+)_{2 3}=\af' \; ,\; (V_-)_{2 1}=-\af' \; ,\;
(\ov{V}_+)_{1 2}=\gh' \; ,\; (\ov{V}_-)_{3 2}=-\gh' \; ,
\label{v33}
\ear
where only the non-vanishing elements of the odd generators are given.
The multiplets normalizations $\af'$ and $\gh'$ satisfy
\beq
4 \af' \gh' = [2] \; .
\eq

The corresponding $R$-matrix, denoted by $r$, is found using methods
similar to those used to determine the $R$-matrix of the four-dimensional
representation. The matrix $r$ depends only on
the product $\af' \gh'$, and no choice of $\af'$ and $\gh'$ is necessary.
The non-vanishing elements of $r$ are given by:
\bars
r &=& 1 \; \; {\rm for} \; \; B_1\otimes B_1 \; ,\; B_2\otimes B_2 \; , \\
r &=& \frac{xq^4-1}{x-q^4}  \; \; {\rm for} \; \; F\otimes F \; , \\
r &=& \left( \begin{array}{cc}
j & i \\
h & j \end{array} \right) \;  {\rm for \; the \; bases}\; (B_1\otimes F,
F\otimes B_1) , \\
& & (B_1\otimes B_2,B_2\otimes B_1)  ,
\; (F\otimes B_2,B_2\otimes F)  ,
\ears
where
\beq
h(x,q)= \frac{1-q^4}{x-q^4} \; ,\; \; i(x,q)=x h(x,q) \; ,\; \;
j(x,q)=q^2 \frac{x-1}{x-q^4} \; .
\label{jgh}
\eq

\subsection{The $R$-matrix for $[\de]_+\ot [b,\de]$ \label{afco}}

The determination of the `hybrid' $R$-matrix, corresponding to a left
representation $[\de]_+$ and a right representation $[b,\de]$, proceeds
differently. Because the representations are different, the commutation of
the coproduct $\Delta$ with an $\check{R}$-matrix does not hold. Instead
one can still use the universal spectral parameter-independent $R$-matrix
with the two foregoing representations to obtain the form of the matrix and
hence simplify the following calculation. Jimbo has shown \cite{jimbo},
for affine $q$-deformed Lie algebras, that a solution of
\beq
\ov{\hat{\dd}}(g)R(x)=R(x)\hat{\dd} (g) \; ,
\eq
for the generators $g$ of the {\it affine} $q$-deformed algebra
$U_q \widehat{osp(2,2)}$,
automatically satisfies the Yang-Baxter equation. This results was generalized
in \cite{bgz} for superalgebras. The coproducts $\ov{\hat{\dd}}$ and
$\hat{\dd}$ extend the coproducts defined earlier to the affine algebra
$U_q \widehat{osp(2,2)}$. The affine coproducts coincide with the usual ones
for all the generators of $U_q osp(2,2)$ except for the additional three
generators belonging to the affine algebra.
The matrix $R^{3\ot 4}(x,q,b)$
has the same block-diagonal structure as the corresponding spectral
parameter independent matrix. More precisely the matrix acts non-trivially
in the same subspaces. I denote $R^{3\ot 4}$ by $R$ in this section and in
the following sections.
It is enough to look for a solution of the
following equations:
\bar
\label{aff1}
R(x) \dd (V_+) &=& \ov{\dd}(V_+) R(x) \; ,\\
R(x) \dd (\ov{V}_+) &=& \ov{\dd}(\ov{V}_+) R(x) \; , \\
R(x) [e_0 \ot q^{\frac{h_0}{2}} + x q^{-\frac{h_0}{2}}\ot e_0] &=&
[e_0 \ot q^{-\frac{h_0}{2}} + x q^{\frac{h_0}{2}}\ot e_0] R(x) \; ,
\label{aff}
\ear
where $h_0=4 S_3$ and $e_0$ is proportional to the $q$-deformed generator
$S_-$ (the normalization is irrelevant). With $f_0$ proportional to the
$q$-deformed generator $S_+$, the three generators $e_0$, $f_0$ and $h_0$
correspond to the additional root of the $q$-deformed affine superalgebra
$U_q \widehat{osp(2,2)}$. An equation similar to (\ref{aff}) holds for
$f_0$.
Note that, for the representations at hand, the generators on the left
of the tensor products
are in the three-dimensional representation while those on the right are
in the four-dimensional one. This system of equations is overdetermined in
the twenty two unknowns of $R(x)$. This is always the case for such systems.
I take $\af=\gh$, $\be=\ep$ in (\ref{vrep1}--\ref{vrep2}), and $\af'=\gh'$
in (\ref{v33}).
Solving the linear equations (\ref{aff1}--\ref{aff}), I
obtain the non-vanishing
elements of the $12 \times 12$ $R$-matrix:
\bar
R &=& q^{1-2b} \; \; {\rm for } \; \; B_1\ot b_1 \; ,\; B_1\ot f_1 \; ,
\; B_2\ot f_1 \; ,\; B_2\ot b_2 \; , \nonumber \\
R &=&\frac{x q^4-q^{1-2b}}{x-q^{3+2b}} \;\; {\rm for} \;\; F\ot f_2 \; ,
\nonumber \\
R &=&\left( \begin{array}{cc}
\frac{x q^{3-2b}-q^2}{x-q^{3+2b}} & -\frac{q^{\frac{5}{2}-b} [1]_+^{1/2}
[1]_-^{1/2}[1+2b]_-^{1/2}x}{x-q^{3+2b}} \\
-\frac{q^{\frac{5}{2}-b} [1]_+^{1/2}
[1]_-^{1/2}[1+2b]_-^{1/2}}{x-q^{3+2b}}
& \frac{x q^2-q^{3-2b}}{x-q^{3+2b}}
\end{array} \right)  \nonumber
\ear
for the basis $(B_1\ot f_2, F\ot b_1)$,
\bar
R &=&\left( \begin{array}{cc}
\frac{x q^2-q^{3-2b}}{x-q^{3+2b}} & -\frac{q^{\frac{5}{2}-b} [1]_+^{1/2}
[1]_-^{1/2}[1+2b]_-^{1/2}x}{x-q^{3+2b}} \\
-\frac{q^{\frac{5}{2}-b} [1]_+^{1/2}
[1]_-^{1/2}[1+2b]_-^{1/2}}{x-q^{3+2b}}
& \frac{x q^{3-2b}-q^2}{x-q^{3+2b}}
\end{array} \right)  \nonumber
\ear
for the  basis $(F\ot b_2, B_2\ot f_2)$,
\bar
R &=& \left( \begin{array}{ccc}
\frac{x q^{3-2b}-q^2}{x-q^{3+2b}} &  \frac{q^{\frac{3}{2}-b}[1-2b]_-^{1/2}
[1]_+^{1/2} [1]_-^{1/2}x}{x-q^{3+2b}}
& -\frac{x q^2 [2]_-}{x-q^{3+2b}} \\
-\frac{q^{\frac{7}{2}-b} [1-2b]_-^{1/2}[1]_+^{1/2}
[1]_-^{1/2}}{x-q^{3+2b}} & \frac{x- q^{5-2b}}{x-q^{3+2b}} &
\frac{q^{\frac{3}{2}-b}[1-2b]_-^{1/2}
[1]_+^{1/2} [1]_-^{1/2}x}{x-q^{3+2b}} \\
-\frac{q^{3-2b} [2]_-}{x-q^{3+2b}} &
-\frac{q^{\frac{7}{2}-b} [1-2b]_-^{1/2} [1]_+^{1/2}
[1]_-^{1/2}} {x-q^{3+2b}} & \frac{x q^{3-2b}-q^2}{x-q^{3+2b}}
\end{array} \right) \nonumber
\ear
for the basis $(B_1\ot b_2, F\ot f_1, B_2\ot b_1)$.
The matrix $R^{3\ot 4}$  has the following properties:
\beq
R(0,q,b)=R(q,b) \; ,\; R^{-1}(x,q,b)= q^{4b-2} R^{ST}(x^{-1},q,b) \; ,
\eq
where $R(q,b)$ is the spectral parameter-independent $R$-matrix  and
the superscript $ST$ denotes a supertransposition for the
non-positive definite scalar product.

\subsection{A Fusion Equation}

The next step in obtaining Bethe ansatz equations, consists of obtaining
a fusion equation that will give a functional relation between
the eigenvalues of the four-dimensional model and those of the hybrid model.
First one notices that the three-dimensional $R$-matrix becomes
block-diagonally proportional to a five-dimensional projector for $x=q^{-4}$.
Define a projector $P$ as
\beq
P=A \: r(q^{-4},q) \;\; {\rm with} \;\;
A={\rm diag}(1,\frac{a}{2},\frac{a}{2},\frac{a}{2},1,\frac{a}{2},
\frac{a}{2},\frac{a}{2},1) \;\; {\rm where} \;\; a=q^2+q^{-2} \; .
\eq
Let $p\equiv 1-P$. One can now do the fusion in the auxiliary space.
Define the matrix
\beq
\tilde{R}_{\langle 1 2 \rangle 3}(x,q,b)=B_{1 2} p_{1 2} R_{1 3}(x q^{-2},q,b)
R_{2 3}(xq^2,q,b) p_{1 2}B^{-1}_{1 2} \; ,
\label{fus}
\eq
where the nine-dimensional matrix $B$ is the change of basis matrix, to the
basis that diagonalizes both projectors, $P$ and $p$. $B$ is given
by three ones and three blocks
\beq
\left( \begin{array}{cc}
-q^2 & 1 \\
q^2 & 1 \end{array} \right) \;.
\eq

The matrix $\tilde{R}$ has $36^2$ elements. However there are many rows and
columns of zeroes. One can  remove them and obtain a $16\times 16$ $R$-matrix
for $[\tde,\de]\ot [b,\de]$.
The matrix (\ref{fus}), which was denoted $R^{4'\ot 4}$ in section
(\ref{philo}),  satisfies the various Yang-Baxter equations discussed in the
same section.

The supertrace, in the spaces 1 and 2, of eq. (\ref{fus}) replicated to
an $L$-site chain, gives a relation between
the transfer matrices of the $[\de]_+\ot [b,\de]$, $[\tde,\de]\ot [b,\de]$
and $[1]_+\ot [b,\de]$ lattices:
\beq
-\tau^{(4'4)}(x)=\tau^{(34)} (x q^{-2}) \tau^{(34)} (x q^2) + \tau^{(54)}(x)
\; .
\label{fus1}
\eq
The superscripts refer to the dimensions of the representations.
The additional minus sign accounts for a mismatch in the grading of the bases.
As explained in section (\ref{philo}), all the transfer matrices in
(\ref{fus1}) commute for all values of the
spectral parameter $x$.
Therefore  the same relation holds for the  respective eigenvalues.

\subsection{Bethe Ansatz \label{bean}}

I now perform a nested algebraic Bethe
ansatz calculation to diagonalize the transfer matrix of
the the $3\ot 4$-model (see \cite{essl} for instance).
The matrix $L_n$ defined above is given by (\ref{l}) and $R^{3\ot 4}(x)$. In
component $L_n$ reads
\beq
L(u)_{a \alpha ,b  \beta}= R^{3\ot 4}(u)_{a \alpha ,b  \beta} \; , \;
1 \leq a,b \leq 3 \; ,\; 1\leq \alpha ,\beta \leq 4 \; ,
\label{l34}
\eq
where the Latin indices correspond to the auxiliary
space, and the Greek indices to the quantum space.
The monodromy matrix (\ref{mono})
satisfies (as explained in section (\ref{philo})):
\beq
\check{r}_{1 2}(u-v) T(u) \ot T(v)=
T(v) \ot T(u)\check{r}_{1 2}(u-v) \; .
\label{rtt1}
\eq
In components, with all Latin indices varying from 1 to 3, and repeated
indices indicating summations, (\ref{rtt1}) reads:
\bar
\lefteqn{\check{r}_{d_1 d_2, b_1 b_2}(u-v) T_{b_1 c_1}(u)
T_{b_2 c_2}(v) (-1)^{\varepsilon_{b_2}
(\varepsilon_{b_1} + \varepsilon_{c_1})} =} \nonumber \\
& &\hspace{-0.4cm} T_{d_1 b_1}(v) T_{d_2 b_2}(u)
\check{r}_{b_1 b_2, c_1 c_2}(u-v)
(-1)^{\varepsilon_{d_2} (\varepsilon_{d_1} + \varepsilon_{b_1})} \; , \;\;
\varepsilon_1=\varepsilon_3=0 \; ,\; \varepsilon_2=1 \; .
\label{e69}
\ear
The monodromy matrix (\ref{mono}), in components, is given by:
\bar
(T(u)^{a b})_{\alpha_1,...,\alpha_L;\beta_1,...,\beta_L} &=&
L(u)_{a\alpha_L , c_L\beta_L}L(u)_{c_L \alpha_{L-1},c_{L-1}\beta_{L-1}}\cdots
\nonumber \\
& &\times L(u)_{c_2 \alpha_1 , b \beta_1}
(-1)^{\sum_{j=2}^L (\varepsilon_{\alpha_j}+
\varepsilon_{\beta_j})\sum_{i=1}^{j-1}\varepsilon_{\alpha_i}} \; .
\label{mono1}
\ear
The signs arise because of the graded tensor product.

There are two obvious candidate for the `highest weight vector', or
ferromagnetic vacuum,
$|\Omega_1\rangle = f_1^{(1)}\ot ... \ot f_1^{(L)}$ and
$|\Omega_2\rangle = f_2^{(1)}\ot ... \ot f_2^{(L)}$, where $f_1$ and $f_2$
were defined in (\ref{basis}).
The two vectors constructed out of $b_1$, or $b_2$, turn out not to
be annihilated by enough $T_{i j}$'s to perform the ansatz. In what follows
I give the results for $|\Omega_1\rangle$.
I shall use the following notation for the matrix of operators $T$:
\beq
T= \left( \begin{array}{ccc}
D_{1 1} & C_1 & D_{1 3} \\
B_1 & A & B_3 \\
D_{3 1} & C_3 & D_{3 3} \end{array} \right) \; .
\eq
The dependence on $u$, or $x=e^{iu}$, is implicit.
The action of $T$ on $|\Omega_1\rangle$ can be summarized in
\beq
T |\Omega_1 \rangle = \left( \begin{array}{ccc}
(R_{2 2})^L & C_1 & 0 \\
0 & (R_{6 6})^L & 0 \\
0 & C_3 & (R_{10\, 10})^L \end{array} \right) |\Omega_1 \rangle \;  .
\label{vac}
\eq
Recall that the $R$-matrix elements are those of $R^{3\ot 4}(x)$.
The fermionic `creation' operators are given by $C_1=T_{1 2}$ and
$C_3=T_{3 2}$. These operators change the spin of the vector they
are acting on by units of $-\de$ and $\de$, respectively.
The eigenvector ansatz is
\beq
|\underline{u}, F\rangle= C_{a_1}(u_1) C_{a_2}(u_2) ... C_{a_n}(u_n)
|\Omega_1\rangle F^{a_n ... a_2 a_1} \; ,\;  a_i \in \{1,3\} \; ,
\eq
where the spectral parameters $\underline{u}$ and the coefficients $F$ are
to be determined. The transfer matrix is
\beq
\tau={\rm Str}_{\rm a} (T)=T_{1 1}+T_{3 3}-T_{2 2} \; .
\eq
Note that a twist on the periodic boundary conditions can be introduced
at this level.

The calculation now proceeds along the usual lines
of the nested algebraic Bethe ansatz method. One pushes
the three pieces of the transfer matrix through the creation operators of the
eigenvector, and obtains `wanted' and `unwanted' contributions. The former
contributions
are proportional to the eigenvector ansatz, while the latter are forced to
vanish, giving a condition on the coefficients $F$. More precisely
$F$ has to be  an eigenvector of a spin-chain on $n$ sites, with
inhomogeneities given by the $u_i$'s,  and constructed
out of a $4\times 4$ $R$-matrix obtained from the nine-dimensional
matrix $r$. By doing a similar ansatz,
one obtains the eigenvectors of this chain, with one set of equations.
Requiring $F$ to be an eigenvector gives another set of equations.

\subsection{Eigenvalues and Bethe Ansatz Equations}

The eigenvalues of the `hybrid' model
can be written as:
\bar
\Lambda^{3\ot 4}(x) &=& (R_{2 2})^L \prod^{m}_{j=1}\frac{1}{j(y_j x^{-1})}
+(R_{2 2})^L \prod_{i=1}^n \prod_{j=1}^m \frac{1}{j(x_i x^{-1})
j(x y_j^{-1})} \nonumber \\
& &-(R_{6 6}(x))^L \prod_{i=1}^n \left( \frac{-g(x x_i^{-1})}
{j(x x_i^{-1})}\right) \, ,
\label{ev}
\ear
where the parameters $x_i$ and $y_j$ are solutions of the Bethe
ansatz equations
\bar
\label{bae11}
\left(\frac{R_{6 6}(x_k)}{R_{2 2}} \right)^L &=&
\prod_{j=1}^m \frac{1}{j(x_k y_j^{-1})} \; , \; i=1,...,n \; ,\\
\prod_{i=1}^n j(x_i y_l^{-1}) &=& \prod_{\stackrel{j=1}{j\neq l}}^m
\frac{j(y_j y_l^{-1})}{j(y_l y_j^{-1})} \; , \; l=1,...,m \; .
\label{bae12}
\ear
These Bethe ansatz equations reflect the choice of a particular grading of
the bases.

The $R$-matrix of the lattice model with auxiliary space $[\frac{3}{2},
\de]$ and quantum space $[b,\de]$ and the eigenvalues of the corresponding
transfer matrix are obtained from the fusion eq. (\ref{fus1}).
I find for the eigenvalues (up to an unessential factor $q^{(2-4 b)L}$):
\bar
\Lambda^{4'\ot 4}(x) &=& \prod_{i=1}^n
\frac{x_i q^{-3}- x q^3}{x_i q^{-1}-x q}
-\left(\frac{x q^{-3/2 + b} - q^{3/2- b}}{x q^{-1/2 -b}-q^{1/2+ b}}\right)^L
\nonumber\\
& & \times \left( \prod_{i=1}^n \prod_{j=1}^m
\frac{x q -y_j q^{-1}}{x q^{-1}-y_j q}
\frac{x q^3 -x_i q^{-3}}{x q -x_i q^{-1}} + \prod_{i=1}^n \prod_{j=1}^m
\frac{x q^{-3}-y_j q^3}{x q^{-1}-y_j q}
\frac{x q^3-x_i q^{-3}}{x q^{-1}-x_i q} \right)\nonumber\\
& &+\left(\frac{x q^{b-3/2}-q^{3/2-b}}{x q^{-1/2-b}-q^{b+1/2}}
\frac{x q^{b-7/2}-q^{7/2-b}}{x q^{-5/2-b}-q^{5/2+b}}\right)^L
\prod_{i=1}^n \frac{x q^3- x_i q^{-3}}{x q^{-1}-x_i q} \; .
\label{eve}
\ear
Recall that $4'\equiv [\tde,\de]$ (the auxiliary space) and $4\equiv [b,\de]$
(the quantum space).
Both eigenvalues (\ref{ev}) and (\ref{eve})  seem to have poles at values
related to the ansatz parameters, $x_i$ and $y_j$. However, the Bethe ansatz
equations ensure
precisely that the residues, at these apparent poles, vanish. This is
generally the case with a Bethe ansatz solution. It is also possible to obtain
from (\ref{fus1}) the eigenvalues of the transfer matrix
based on the $R$-matrix of the representation $[1]_+\ot [b,\de]$.

I now rewrite the Bethe ansatz
equations, (\ref{bae11}) and (\ref{bae12}), in the `generic form'.
For $x=e^{i u}$, $q=e^{i \gamma /2}$ and some redefinitions,
the equations become:\footnote{This $\gamma$ should not be confused with
the one entering the generators of $osp(2,2)$ for the representation $[b,
\frac{1}{2}]$.}
\bar
\label{bae21}
\left( \frac{\sinh \frac{1}{2}(u_k - (b-1/2) i \gamma)}
{\sinh \frac{1}{2}(u_k + (b-1/2) i \gamma)}\right)^L =
\prod_{j=1}^m \frac{\sinh \frac{1}{2}(u_k -v_j + i \gamma)}
{\sinh \frac{1}{2}(u_k -v_j - i \gamma)} \; ,\;\;
1\leq k \leq n \; , \\
\prod_{k=1}^n \frac{\sinh \frac{1}{2}(u_k -v_j + i \gamma)}
{\sinh \frac{1}{2}(u_k -v_j - i \gamma)}= -\prod_{l=1}^m
\frac{\sinh \frac{1}{2}(v_l -v_j +2 i \gamma)}
{\sinh \frac{1}{2}(v_l -v_j -2 i \gamma)}\; , \;\; 1\leq j\leq m \; .
\label{bae22}
\ear

The choice of the ferromagnetic vacuum $|\Omega_2\rangle$ for the eigenvector
ansatz gives a different set of equations. They differ from
(\ref{bae21}--\ref{bae22}) by the change $b-\de \rightarrow -(b+\de)$.
The eigenvalues also differ in this manner. Such a difference between
the Bethe ansatz results indicate that the eigenvectors obtained from
both $|\Omega_1\rangle$ and $|\Omega_2\rangle$ may be necessary
to have a complete set
of eigenvectors. This will be investigated elsewhere.

\subsection{A Conjecture}

As indicated earlier, the representation $[b,\de]$ cannot be obtained from a
tensor product of smaller representations (if $b\neq \pm\tde$).
I conjecture
the result for the eigenvalues of $\tau^{(44)}$ transfer matrix,
based on the Bethe ansatz
equations  already obtained, and
which also hold for $4\ot 4$ since they depend only on the quantum space
and the algebra, and other considerations. More precisely, I replace the
prefactors raised to the power $L$ in (\ref{eve}),
by the ones corresponding to the matrix
$\tau^{4\ot 4}$, acting on the eigenvector $|\Omega_1\rangle$.
Then this eigenvalue
is `dressed' by products obtained through a
`minimal' modification of the terms inside the products of the eigenvalue
(\ref{eve}). This modification must preserve the Bethe ansatz equations
found earlier, meaning, the vanishing of the residues at the apparent poles
must yield the same equations. Finally the eigenvalues of the Hamiltonian
must be generically real.
The eigenvalues are then given by:
\bar
\lefteqn{\Lambda^{4\ot 4}(u)= \left(\frac{\sin \frac{1}{2}(u + (1-2 b)\gamma)}
{\sin \frac{1}{2}(u - (1-2 b)\gamma)}\right)^L
\prod_{k=1}^n \frac{\sinh \frac{1}{2}(u i + u_k + (b-\frac{1}{2})\gamma i)}
{\sinh \frac{1}{2}(u i + u_k - (b - \frac{1}{2})\gamma i)}} \nonumber \\
& &-\left(\frac{\sin(\frac{u}{2})}{\sin
\frac{1}{2}(u - (1 - 2 b)\gamma)}\right)^L
\left(\prod_{k=1}^n\prod_{l=1}^m\frac{\sinh\frac{1}{2}(u i + v_l +
(\frac{3}{2}-b)\gamma i)}{\sinh \frac{1}{2}(u i + v_l -
(\frac{1}{2}+b)\gamma i)} \right. \nonumber\\
& & \times \frac{\sinh \frac{1}{2}(u i + u_k +
(b-\frac{1}{2})\gamma i)} {\sinh \frac{1}{2}(u i + u_k -
(b - \frac{1}{2})\gamma i)}
+ \prod_{k=1}^n\prod_{l=1}^m
\frac{\sinh\frac{1}{2}(u i + v_l - (\frac{5}{2}+b)\gamma i)}{\sinh
\frac{1}{2}(u i + v_l - (\frac{1}{2}+b)\gamma i)} \nonumber\\
& & \left. \times \frac{\sinh \frac{1}{2}
(u i + u_k + (b-\frac{1}{2})\gamma i)} {\sinh \frac{1}{2}(u i + u_k -
(\frac{3}{2}+b)\gamma i)} \right)
+\left(\frac{\sin(\frac{u}{2})
\sin\frac{1}{2}(u-2\gamma )}{\sin\frac{1}{2}(u-(1+2 b)\gamma)
\sin\frac{1}{2}(u-(1-2 b)\gamma)}\right)^L\nonumber\\
& & \times \prod_{k=1}^n \frac{\sinh \frac{1}{2} (u i + u_k + (b-\frac{1}{2})
\gamma i)}{\sinh \frac{1}{2}(u i + u_k - (\frac{3}{2}+b)\gamma i)}\; .
\label{ev44}
\ear
The integers $n$ and $m$ can be restricted to
\beq
0 \leq m \leq n \leq 3 L \; ;
\eq
the ansatz eigenvectors would otherwise identically vanish.
This follows from the analysis of the action of the creation operators on
the generating vectors at both levels of the Bethe ansatz.

Considering
higher dimensional representations, and the relations between
the respective transfer matrices, one can obtain functional relations for the
eigenvalues of the transfer matrices, including the transfer matrix
of the lattice with both, auxiliary and quantum spaces, equal to $[b,\de]$.
These relations can serve as a check for (\ref{ev44}).

\subsection{Hamiltonian and Momentum}

The logarithmic derivative of the transfer matrix $\frac{1}{i}\tau^{4\ot4}$ at
$x=1$ or $u=0$ gives a spin-chain  Hamiltonian. This Hamiltonian has
coupling constants that depend on $\gamma$ and $b$. However the eigenvalues
can be real for certain excitations. Dropping the contribution from the
prefactor, which amounts to a translation of the energy origin, the
eigenvalues of the Hamiltonian are given by
\beq
E=\de \sum_{k=1}^n \frac{\sin ((\frac{1}{2}-b)\gamma)}
{\sinh \frac{1}{2} (u_k + (\frac{1}{2}-b)\gamma i)
\sinh \frac{1}{2} (u_k - (\frac{1}{2}-b)\gamma i)} \; .
\label{ener}
\eq

The eigenvalues of the momentum operator, $P=\frac{1}{i}\ln \tau(u=0)$,
are given by
\beq
P=\frac{1}{i}\sum_{k=1}^n \ln \left( \frac{\sinh\frac{1}{2}(u_k + \gamma
(b-\frac{1}{2}) i)}{\sinh \frac{1}{2} (u_k - \gamma (b-\frac{1}{2})i)}\right)
\; ,
\label{momen}
\eq
up to a translation by a constant.

\subsection{The Parameter $b$: a Generalized spin}

The Bethe ansatz equations (\ref{bae21}) and (\ref{bae22})  have the
form expected from the algebra and the highest weight label of the
representation $[b,\de]$. The $b-\de$ appearing in the left-hand-side of
(\ref{bae21}) is proportional to the scalar product of this highest weight
and the first root of the simple root basis.

The comparison of the Bethe ansatz equations, (\ref{bae21}) and
(\ref{bae22}), with those of the $SU(2)$-chain with arbitrary
spin, shows why $b-\de$ can be considered as some equivalent of a
continuous spin label.
In contrast to the generalized spin of \cite{gs}, which
only exist for $q$ a root of unity, the `spin' $b-\de$ exist for
all values of $q$.

For $b$ in certain ranges with rational bounds, the value
of the central charge is independent of $b$; however the values
of the central charge will be different for each domain. The conformal weights
will depend on $b$ continuously throughout each domain.

\section{Conclusion}

The motivation for studying lattice models with an underlying
$U_q osp(2,2)$ symmetry arose from an attempt at constructing lattice
models which in a certain continuum limit could yield $N=2$ superconformal
field theories. The supersymmetry may then be traced back to the lattice.
The first step in such an approach consists of determining $R$-matrices.
In this paper I have derived trigonometric $R$-matrices which depend
on two continuous generic  parameters $q$ and $b$.  The origin of the second
parameter is a $U(1)$ generator in the Cartan subalgebra.\footnote{$R$-matrices
with  a double parametric dependence have been derived
for $U_q su(2)$ \cite{gs}; however
there the deformation parameter $q$ has to be a root of unity.}
I give the rational limits of the matrices obtained; they also depend
on the parameter $b$.
A Bethe ansatz diagonalization of the transfer matrix
is complicated by the representation theory of the superalgebra. After
obtaining the BAE's for a specific model using fusion  I conjecture the
result for the eigenvalues of the $[b,\de]\ot [b,\de]$ lattice. These
results can be readily generalized to twisted periodic boundary conditions.
Preliminary results indicate that the central charge in the continuum limit
does not depend continuously on  $b$ (the conformal weights do however).
These results depend on a careful analysis of the Bethe ansatz equations.
The determination of the vacuum of the model is complicated by the grading
of the algebra.\footnote{In this respect the results I obtain here seem
to contradict the conclusions of \cite{sunm}.} A numerical study will
be useful to find the ground state.
Transfer matrices with periodic boundary conditions as studied here commute
with each other at different values of the spectral parameter, but they do not
commute with the generators of $U_q osp(2,2)$. This does not mean however
that such a symmetry is not present. For $SU(2)$ the underlying
symmetry was exhibited in \cite{saleur} for twisted boundary conditions.
This can presumably be generalized. However by considering open boundary
conditions it is possible to obtain transfer matrices which commute
with the $q$-deformed algebra. To obtain such matrices one needs
to find $K$-matrices. This is part of a work in progress.

\vspace{1cm}
\hspace{-6mm}\Large\bf Appendix \\
\normalsize
\\
I  briefly discuss the rational
limits \cite{aratio} obtained from (\ref{rx}).
Rational limits can be obtained by letting
both $x$ and $q$ tend to one, with $x$ behaving as some power of $q$.
Consider for instance the limit obtained for the choices $\la=q^{1-4 b^2}$,
$q=e^{-i \gh}$ and $x=e^{4 i u\gh}$, where $\gh$ tends to zero.
The resulting rational $R$-matrix is
\beq
\check{R}(u,b)=P_1(q=1,b)+g_2(u,b) P_2(q=1,b) +g_3(u,b) P_3(q=1,b)\;,
\label{rub}
\eq
where
\beq
g_2(u,b)=\frac{1+2 b - 2 u}{1+2 b +2 u} \;\; , \; \;
g_3(u,b)=\frac{1-2 b - 2 u}{1-2 b +2 u} \; .
\eq

The block-diagonal matrix (\ref{rub}) contains:

(i) one-dimensional sub-matrices,
\bar
\check{R}(u,b) &=& 1 \;\; {\rm for} \;\; b_i\otimes b_i \; ,
\; i=1,2 \; , \\
\check{R}(u,b) &=& g_2(u,b) \;\; {\rm for} \;\; f_2\otimes f_2 \; , \\
\check{R}(u,b) &=& g_3(u,b) \;\; {\rm for} \;\; f_1\otimes f_1 \; ,
\ear

(ii) two-dimensional sub-matrices,
\beq
\check{R}(u,b)=\frac{1}{1-2b+2u} \left(
\begin{array}{cc}
1-2b & 2u \\
2u & 1-2b \end{array} \right) \; ,\eq
for the two bases $(b_1\otimes f_1, f_1\otimes b_1)$ and
$(f_1\otimes b_2, b_2\otimes f_1)$,
\beq
\check{R}(u,b)=\frac{1}{1+2b+2u} \left(
\begin{array}{cc}
1+2b & 2u \\
2u & 1+2b \end{array} \right) \; ,
\eq
for the two bases $(b_1\otimes f_2, f_2\otimes b_1)$
and $(f_2\otimes b_2,b_2\otimes f_2)$,

(iii) a four-dimensional sub-matrix,
\bar
\check{R}(u,b)=\frac{1}{d(u,b)}\left(
\begin{array}{cccc}
1-4b^2+4u & n(u,b) & n(u,b) & 4u^2 \\
-n(u,b) & 1-4b^2 & -4u(1+u) & n(u,b)  \\
-n(u,b) & -4u(1+u) & 1-4b^2 & n(u,b)  \\
4u^2 & -n(u,b) & -n(u,b) & 1-4b^2+4u
\end{array} \right) \, ,
\label{rfour}
\ear
where
\beq
n(u,b)=2u(1+2b)^{1/2}(1-2b)^{1/2} \;  ,\; d(u,b)=(1+2b+2u)(1-2b+2u) \; ,
\eq
for the basis $(b_1\otimes b_2,f_1\otimes f_2,f_2\otimes f_1,b_2\otimes b_1)$.
The rational matrices have the block structure of their trigonometric
parents, and they satisfy a Yang-Baxter equation with $u$ as an additive
spectral parameter.
The parameter $b$ appears explicitly in (\ref{rub}); it cannot be scaled away
by rescaling the spectral parameter $u$.

\vspace{1cm}
\hspace{-6mm}\Large\bf Acknowledgements \\
\normalsize
\\
I would like to thank H. Saleur for numerous discussions
and N. Warner for continued support. I am grateful to J. Sch\"{u}lze
for lively and educating discussions on quantum groups.

\end{document}